\documentstyle[equations,11pt]{article}

\renewcommand{\appendix}
        {
        \par
        \setcounter{section}{0}
        \setcounter{subsection}{0}
        \gdef\afterthesectionpunctdefault{:}
        \gdef\thesection{{Appendix \Alph{section}}}
        \renewcommand{\theequation}{\Alph{section}\arabic{equation}}
        \setcounter{equation}{0}
        }
\def\lsim{\hbox{\lower .8ex\hbox{$\, \buildrel < \over \sim\,$}}}
\def\gsim{\hbox{\lower .8ex\hbox{$\, \buildrel > \over \sim\,$}}}
\def\ds{\displaystyle}
\newcommand{\bit}{\begin{itemize}}
\newcommand{\eit}{\end{itemize}}

\newcommand{\be}{\begin{equation}}
\newcommand{\ee}{\end{equation}}
\begin{document}
\begin{center}
{\Large{\bf Bound states of $^3$He in $^3$He--$^4$He mixture films}}

\vspace{1.0cm}

{\large Eugene Bashkin$^{1,2}$, Nicolas Pavloff$^3$
and Jacques Treiner$^3$ }
\end{center}

\vspace{1.0 cm}

\noindent $^1$Department of Physics and Material Sciences Center, Philipps
University, 35032 Marburg, Fed. Rep. Germany \hfill\break
\noindent $^2$Kapitza Institute for Physical Problems, 117334 Moscow,
Russia \hfill\break
\noindent $^3$Division de Physique Th\'eorique\footnote { Unit\'e de
Recherche des Universit\'es Paris XI et Paris VI associ\'ee au CNRS.},
Institut de Physique Nucl\'eaire, F-91406 Orsay Cedex, France \hfill\break

\vspace{1.5 cm}
\begin{center}
{\bf abstract}
\end{center}

	$^3$He atoms dissolved in superfluid $^4$He may form dimers
($^3$He)$_2$ in two-dimensional (2D) geometries. We study dimer
formation in films of dilute $^3$He--$^4$He mixture. After designing a
schematic $^3$He--$^3$He interaction potential we calculate the dimer
binding energy for various substrates. It is shown that $^3$He impurity
states localized near the substrate give rise to the largest
magnitudes of the binding energies.

\vspace{3cm}
\noindent PACS numbers :\hfill\break
\noindent 67.70.+n ~Helium films.\hfill\break
\noindent 67.60.Fp ~He II -- $^3$He mixtures.\hfill\break

\vspace{2cm}
\noindent IPNO/TH 94-56 \hspace{0.5cm} {\it to appear in J. Low Temp.
 Phys.}\hfill\break
\newpage

\section{Introduction}

	$^3$He atoms dissolved in bulk $^4$He have always been considered an
ideal system for testing the Landau approach describing the macroscopic
properties of quantum fluids in terms of elementary excitations
(quasiparticles) \cite{1}. At low enough temperatures one can neglect the
contribution of the excitations pertaining to superfluid $^4$He (phonons and
rotons). The system is then described in terms of an interacting gas of $^3$He
quasiparticles imbedded in the uniform background of superfluid $^4$He. A bare
quasiparticle has a particle--like energy spectrum
\cite{pom} with an effective mass larger than the bare mass. The quasiparticle
interaction at long distances is the bare $^3$He--$^3$He interatomic
interaction renormalized by a phonon--induced term \cite{2}. At small distances
a strong repulsive core is predominant. However, the overall effect turns out
to be attractive, as was demonstrated on the basis of theoretical calculations
\cite{3,4} and experimental data \cite{5,6} on the s--wave scattering length
$a$ ($a<0$).

	Despite the fact that the effective attraction between two $^3$He
quasiparticles is too weak to lead to a bound state in the bulk, such a bound
state -- a ($^3$He)$_2$ dimer -- should certainly exist in helium systems with
reduced dimensionality \cite{7,7b}. It is well known that
in one and two dimensions {\it any} attractive potential satisfying the
perturbation theory criterion gives rise to a bound state (see {\it e.g.} Ref.
\cite{8}). The result does not apply, however, in the case of a nonperturbative
potential like the $^3$He--$^3$He interaction, with a strong repulsion at short
distances. Nevertheless, one can rigorously prove that spinless ($^3$He)$_2$
dimer {\it must} exist in quasi--2D and quasi--1D geometries (like films and
narrow capillaries of $^4$He) provided  $a$ is negative and the characteristic
scale of confinement of the $^3$He atoms is much larger than the interaction
range -- which is of the order of $|a|$ -- (see \cite{7b} and appendix A).
When the system is cooled down to temperatures lower than the corresponding
binding energy, the single impurity quasiparticles form ($^3$He)$_2$
Bose--dimers, and the Fermi component of $^3$He is replaced by a new Bose
quantum fluid of ($^3$He)$_2$ of reduced dimensionality. This phenomenon should
strongly affect the macroscopic properties of dilute mixtures resulting in an
extra superfluid transition, new features of the phase diagram, anomalous sound
absorption, {\it etc.} \cite{7}. This is the reason why calculating the binding
energy is the main issue in the theory of dimerized $^3$He--$^4$He solutions.
To be specific we will concentrate on dilute $^3$He--$^4$He mixtures in various
2D geometries ({\it i.e.} in films on various substrates).

	In $^4$He films, $^3$He impurities are localized in the direction
normal to the substrate over a characteristic width $w$ which can be much
smaller than the film thickness. When extrapolated to small $w$ (of the order
of a few atomic layers) the theory of Ref. \cite{7,7b} provides an estimate
for the binding energy which seems to be quite attainable for experimental
studies. However, it is difficult to assess the accuracy of such an estimate
because the theory in question certainly does not hold in such restricted
geometry, {\it i.e.} in a real 2D
situation (when $w \sim a$). In a pure 2D case one cannot {\it a priori} be
sure if a bound state of two $^3$He atoms exists at all. Therefore,
calculating the binding energy for various film thickness
and different substrates may provide direct clue on which experimental
conditions are best suited for the detection of ($^3$He)$_2$ dimers. Such
calculations are the primary goal of this paper. Here we focus on two cases
where $^3$He impurities adopt a 2D configuration :

\bit
\item[(i)] The Andreev states of $^3$He at the free surface of $^4$He (see
\cite{and,9,10} and references therein).

\item[(ii)] The $^3$He states localized at the interface between a $^4$He film
and the substrate, as predicted in \cite{11,11b}.
\eit

	Case (ii) seems particularly favorable for the creation of dimers
because $^3$He quasiparticles near the substrate have a large effective
mass (due to the higher $^4$He density) and a narrow wave-function (localized
within the first $^4$He liquid layer). The localization length (the width of
the wave function in the direction across the film) is of the order of a
few angstr\"oms. In Sec. 2 some possible realizations of a quasi--2D ensemble
of $^3$He impurities in superfluid $^4$He films on various substrates are
reviewed. In Sec. 3 we derive a schematic $^3$He--$^3$He interaction
and then evaluate in Sec. 4 the corresponding ($^3$He)$_2$ binding energy. We
give our concluding remarks in Sec. 5.

\section{$^3$He impurities in $^4$He films.}

	We assume in the following that inhomogeneities of the substrate in the
$x-y$ plane parallel to the surface play a negligible role and we impose
translational invariance on the helium densities ($^3$He and $^4$He). This
does not imply that possible corrugation of the surface is neglected. Indeed,
in that case, $^4$He atoms will certainly fill the dips of the surface until it
is more or less flat ; the effective substrate to be considered is then a
mixture of the original substrate and these helium atoms, the effect of which
is to weaken the potential.

	The first question to address is to determine the state of  liquid
helium on a given substrate. The different situations to expect, as a function
of increasing attractive strength of the substrate potential, are the following
: non wetting, wetting with prewetting \cite{12}, and solidification of one or
two layers.	We limit our study to cases where no solid layer forms near the
substrate : there is a clear qualitative change of $^3$He  impurity states when
$^4$He solidifies ; the $^3$He atoms occupy substitutional states in solid
$^4$He and move through the lattice via tunneling processes (see {\it e.g.}
Ref. \cite{15}). Our continuous description of the solid misses this
phenomenon, since the coupling of $^3$He atoms with the $^4$He matrix is
represented here through a density dependent $^3$He effective mass fitted on
properties of {\it liquid} mixtures (see details in Ref. \cite{11}).

	Hence, in order to specify the validity of our approach we
need a criterion assigning a solid or liquid character to each of the
first $^4$He layers. This has been done in Ref. \cite{13} where different
substrates were characterized by their interaction with helium through two
coefficients $C_3$ and $D$

\be\label{e1}
V_{sub}(z) = {4\over 27 D^2} \left( {C_3\over z^3}\right)^3 -
{C_3\over z^3} \; .
\ee

Here $D$ is the depth of the attractive part of $V_{sub}(z)$ and $C_3$
characterizes the van der Waals tail. According to Ref. \cite{13} one can draw
in the $C_3-D$ plane the lines corresponding to solidification of the first and
second layer (see Fig. 1). For completeness, we have also reproduced, from Ref.
\cite{12}, the line separating the non-wetting region from the wetting one.
Notice that the criterion used in \cite{13} is only approximate and also
that there are large uncertainties in the parameters of the potentials (up to
$\sim$ 30 \% on $D$ \cite{14}). As an extreme case, for hydrogen the value $C_3
= 1000$ K\AA$^3$ extracted from different experiments (see \cite{14b} and
references therein) is significantly larger than the theoretical value of 360
K\AA$^3$ from Ref. \cite{14}. So, the predictions are uncertain for cases close
to a line : it may well be that the first layer does solidify on Mg, and that
the second layer solidifies on MgO, Cu and Ag.

	We now come to the $^3$He impurity states. Calculations were performed,
as in Ref \cite{11}, in the limit of {\it one} $^3$He atom. When
considering finite $^3$He coverage, each state generates a 2D Fermi
disc, characterized by a 2D Fermi momentum. Also, through
self-consistency, the $^4$He profile, the $^3$He average field -- and
consequently the single particle states -- depend on the $^3$He coverage
(see Ref. \cite{10} for a study of finite $^3$He coverage on the $^4$He
bulk surface). We shall neglect these effects in the following
discussion, which is valid for small $^3$He coverage only.

	Variation of the density functional with respect
to the impurity wave function $\phi$ leads to the following equation
\begin{equation}\label{e3}
-\frac{d}{dz}\left( \frac{\hbar^2}{2m^*(z)} \frac{d\phi}{dz} \right)
       + U_{ext}(z) \phi (z) = \epsilon \phi (z) \; ,
\end{equation}

	$U_{ext}(z)$ is the $^3$He mean field, see Ref. \cite{11}. It comprises
a term due to $^3$He--$^4$He interaction plus the substrate potential
$V_{sub}$. The density-dependent effective mass is parametrized as :

\begin{equation}\label{e4}
\frac{\hbar^2}{2m^*({\bf r})} = \frac{\hbar^2}{2m_3}
\left( 1 - \frac{\bar{\rho}_4({\bf r})}{\rho_{4c}} \right)^2 \; ,
\end{equation}

\noindent where $\rho_{4c}=0.062$ \AA$^{-3}$ and $\bar{\rho}_4({\bf r})$ is the
local $^4$He density averaged over a sphere of radius 2.38 \AA~ (see Ref.
\cite{11}). This parametrization is fitted to the pressure dependence
of the $^3$He effective mass in bulk liquid $^4$He \cite{dalstri}.

	Then the effective mass $M^*$ of the $^3$He impurity in a given state
is defined by :
\begin{equation}\label{e5} \frac{\hbar^2}{2M^*} =
  \int_{-\infty}^{+\infty} dz \frac{\hbar^2}{2m^*(z)} \phi^2(z) \;
\end{equation}

	One can check that an energy close to that obtained through
Eq. (\ref{e3}) can be recovered from the effective hamiltonian
containing the constant effective mass $M^*$
\begin{equation}\label{e6}
H = \frac{{\bf p}^2}{2M^*} + U_{ext}(z) \; ,
\end{equation}

\noindent where $U_{ext}$ is the effective $^3$He potential appearing
already in Eq. (\ref{e3}).

	Fig. 2 shows the results for the $^4$He density profile and the
$^3$He wave function near a Cs and a Li surface. Numerical results for
other substrates are displayed in Table 1. The existence of 2D $^3$He
states near a weakly binding surface, although somewhat
counter-intuitive, appears as a general feature of these interfaces.

	The physics behind this feature is best understood by considering a
Lekner approach to the problem. In this variational method, the ansatz
for the wave function $\Psi$ of $N-1$ $^4$He atoms and one $^3$He atom
is taken as
\begin{equation}
\Psi(1,2,...,N) = \phi(1)\, \Psi_0(1,2,...,N)
\end{equation}
where $\phi$ denotes the wave function of the impurity and $\Psi_0$ the
ground state wave function of $N$ $^4$He atoms. Variation of the average
energy of the system with respect to $\phi$ results in a one-body
Schr\"odinger equation (without effective mass effects)
\begin{equation}
-\frac{\hbar^2}{2\, m_3} \, \phi''(z) +U^L_{ext}(z)\, \phi(z) =
                                             \epsilon\, \phi(z)
\end{equation}
where $U^L_{ext}(z)$ is an external field due to both the substrate and the
$^4$He environment,
\begin{equation}\label{eq:Ulek}
U^L_{ext}(z) = \frac{\hbar^2}{2\, m_3}\, \frac{\phi''_4(z)}{\phi_4(z)}
     + \left( \frac{m_4}{m_3} - 1 \right)\, \tau_4(z) + \mu_4 + V_{sub}(z).
\end{equation}
Here $\phi_4(z) = \sqrt{\rho_4(z)}$, $\mu_4$ is the $^4$He chemical
potential, $\tau_4(z)$ the kinetic energy density in the $^4$He ground
state and $V_{sub}(z)$ is the substrate potential.

	In the case of a free surface originally considered by Lekner,
$V_{sub} = 0$; then Eq. (\ref{eq:Ulek}) provides a mechanism for the
formation of the pocket of potential at the surface : the kinetic energy
density $\tau_4$ goes to zero faster than the term in $\phi''_4$. For a
numerical study, one assumes that $\tau_4(z)$ can be parametrized in term
of the local density $\rho_4(z)$. The simple following form was proposed in
Ref. \cite{man79,dal89}

\begin{equation}\label{t4}
\tau_4({\bf r}) = \tau_0\, \left( \frac{\rho_4({\bf r})}{\rho_0} \right)^n -
  \frac{\hbar^2}{2 m_4}
  \frac{\Delta \phi_4(r)}{\phi_4(r)}
\end{equation}

\noindent where the value $\tau_0 = 13.3$ K is chosen  so that the binding
energy of a $^3$He atom in the bulk $^4$He is $-2.8$ K; $\rho_0$ is the
saturation density of liquid $^4$He and $n$ = 1.76 has been related to the
excess volume parameter and the compressibility in dilute $^3$He--$^4$He
mixtures \cite{man79}.

      The effective potential $U^L_{ext}$ resulting from the Lekner theory for
a film of 0.6 \AA$^{-2}$ on a Mg substrate in shown in Fig.3. The average field
$U_{ext}$ obtained using the density functional theory is shown for comparison.
One sees that the Lekner field is slightly more attractive at the surface; the
corresponding energy of the Andreev state is found to be $-5.24$ K (using the
bare $^3$He mass) whereas the density functional result is $-5.27$ K, with $M^*
= 1.34\, m_3$ (the experimental result being $\epsilon = (-5.02 \pm 0.03)$ K
and $M^*/m_3=1.45 \pm 0.1$, see Ref. \cite{Edw78}) . Close to the substrate the
average fields given by the two theories show similar qualitative behaviour,
with attractive wells in correspondence with the oscillations of the $^4$He
density. The Lekner field is more attractive than the density functional one,
and leads to $\epsilon = - 14.1$ K (using the bare mass), while the density
functional theory gives $\epsilon = - 5$ K, with $M^* = 2.9 \, m_3$. We believe
that the density functional results are more reliable, since the model is
fitted to a larger number of bulk mixture properties. The case of a
semi-infinite liquid in contact with a Cs substrate is shown in Fig. 4. The
Lekner value is $\epsilon  = -6.3$ K, compared to the density functional value
$\epsilon = - 4.74$ K, with $M^* = 1.51\, m_3$.

	Note that the Lekner result could be improved by a modification of the
parametrisation (\ref{t4}). Here we only want to point out that substrate
states are not specific to the Density Functional method, but emerge from the
general features of the substrate--helium interface.

 	To summarize, we stress here is that the mechanism by which an
Andreev state is generated at the free surface of liquid helium is also
operating at the liquid substrate interface. On a free surface, a $^3$He
impurity is bound by 5 K (Andreev state), {\it i.e.} by 2.2 K more than in the
bulk. Now a {\it weak} binding substrate {\it perturbs} the energetics of the
free surface in {\it two competitive ways}, namely : i) the wall produces a
readjustment of the $^4$He density profile, which reduces the width of the
$^3$He wave function, and this tends to increase the energy of the state (to
make it less bound); ii) however the attractive substrate potential acts on the
$^3$He atom also and tends to decrease the energy of the $^3$He state. Clearly,
if the perturbation is small enough, the gap of 2.2 K between the unperturbed
surface energy and the bulk energy will not vanish, therefore a bound state
{\it has to} remain at the liquid-wall interface.

	That the argument remains valid also in the case of Mg, which
produces quite a marked layering of the fluid, may seem surprising.
Whether we reach in that case the limit of the model remains to be seen, by
comparing with more microscopic calculations or with experiment.

	As discussed in \cite{11} the existence of a substrate state
allows one understand some unexplained temperature dependence of third
sound velocities reported in Ref. \cite{16}. Several experimental tests
of its existence were proposed in Ref. \cite{11,11b}. Recently,
experimental evidence was reported that $^3$He impurities have a bound
state at the $^4$He liquid-solid interface \cite{noz}, with a binding energy in
fair agreement with the prediction of Ref. \cite{11b}. In the following
we focus on one of the most exciting consequences : the possible formation of
($^3$He)$_2$ dimers with a sizeable binding energy.

\section{ A schematic $^3$He--$^3$He interaction}

	The next step in evaluating the dimer binding
energy is to choose a sensible effective
interaction between the $^3$He quasiparticles. The requirements are three-fold
: (1) the long range attraction is the bare term reduced by a factor $\alpha^2$
where $\alpha$ is the excess volume parameter in $^4$He (see \cite{2,17}
and below); (2) as mentioned in the introduction, the short distance term is
repulsive and equal to the bare potential; (3) the effective potential should
reproduce the s-wave scattering length $a$ of $^3$He in $^4$He ($a \simeq
-0.97$ \AA, see Ref. \cite{5}).

	Let us start with the first requirement ; it will lead us to postulate
a generic form for the effective potential. Our derivation of  the energy of
two $^3$He atoms imbedded in $^4$He is patterned on  the approach of Ref.
\cite{17}. We consider two $^3$He atoms located at points ${\bf r}_1$ and ${\bf
r}_2$. Atom $i$ $(i=1,2)$ occupies a volume $\Omega_3$ centered at ${\bf r}_i$
(we denote it as $\Omega_3[{\bf r}_i]$) of which the $^4$He atoms are expelled
(see Fig. 5). So, if the total volume of the sample is $\Omega$, the $^4$He's
will occupy a volume $\Omega'=\Omega-\Omega_3[{\bf r}_1]- \Omega_3[{\bf r}_2]$.
If $\rho_4$ is the mean $^4$He density, then one can consider that each $^4$He
atom occupies a volume $\Omega_4=1/\rho_4$ and the excess volume parameter is
by definition

\be\label{p0} \alpha = {\ds\Omega_3-\Omega_4\over \ds\Omega_4} \; . \ee

	The total potential energy of the system is
\begin{subequations}
\be\label{p1a} E = E_{33} + E_{34} + E_{44} \; , \ee

\noindent with obvious notations (for instance $E_{34}$ is due to the
$^3$He--$^4$He interaction). $E$ can be separated in a constant ({\it i.e.}
position independent) term plus a term which is an effective $^3$He--$^3$He
potential :

\be\label{p1b} E = C^{\mbox{st}} + V_{eff} (|{\bf r}_1-{\bf r}_2|) \; . \ee
\end{subequations}

	The separation (\ref{p1b}) is made unambiguous by imposing that
$V_{eff}$ goes to zero at large distance (when the two $^3$He impurities are
far apart).

	Let us now study the contributions of $E_{33}$, $E_{34}$ and $E_{44}$
to the effective potential $V_{eff}$.

	If two helium atoms interact {\it via} the bare Lennard-Jones potential
$V_{LJ}$, $E_{33}$ is simply

\be\label{p2} E_{33} = V_{LJ} (|{\bf r}_1-{\bf r}_2|) \; . \ee

\noindent where the explicit form of $V_{LJ}$ is

\be\label{p2b} V_{LJ}(r) = 4 \varepsilon \left[
\left({\ds\sigma\over\ds r}\right)^{12} -
\left({\ds\sigma\over\ds r}\right)^{6} \right] \; ,\ee

\noindent with $\varepsilon = 10.22$ K and $\sigma = 2.556$ \AA.

	The cross term $E_{34}$ is

\be\label{p3} E_{34} =
\int_{\Omega'} V_{LJ}(|{\bf r}_1-{\bf r}|) {d^3r\over\ds\Omega_4} +
\int_{\Omega'} V_{LJ}(|{\bf r}_2-{\bf r}|) {d^3r\over\ds\Omega_4} \; . \ee

	In the r.h.s. of (\ref{p3}) the two integrals are equal. Let us focus
on the first one for instance. The integration domain can be separated
into two
contributions : $\Omega'= (\Omega-\Omega_3[{\bf r}_1])-\Omega_3[{\bf r}_2]$.
The first subdomain  describes the energy due to the introduction of atom $1$
alone in the $^4$He matrix. It corresponds to the "dressing" of the bare
$^3$He particle and plays no role in the quasiparticle interaction ({\it i.e.}
it does not depend on $|{\bf r}_1-{\bf r}_2|$ and can be included in the
constant term of Eq. (\ref{p1b})). The second subdomain describes the
interaction of particle $1$ with fictitious $^4$He atoms occupying
$\Omega_3[{\bf r}_2]$ and brings a contribution to $V_{eff}$, since it
depends of the respective position of the $^3$He impurities. Hence
$E_{34}$ contributes to $V_{eff}(|{\bf r}_1-{\bf r}_2|)$ with the
following term (the factor 2 comes from the sum of the two equal
integrals appearing in (\ref{p3})) :

\be\label{p4} E_{34} \leadsto
-2 \int_{\Omega_3[{\bf r}_2]} V_{LJ}(|{\bf r}_1-{\bf r}|)
{d^3r\over \ds\Omega_4}
\simeq -2{\ds\Omega_3\over\ds\Omega_4} V_{LJ} (|{\bf r}_1-{\bf r}_2|) \; . \ee

	In the last term of (\ref{p4}) we have replaced the integral by an
approximate form valid only if $\vert {\bf r}_1-{\bf r}_2 \vert$ is large
compared with the characteristic radius of $\Omega_3$. If not, the
dressing of the quasiparticles might affect one another (see below).

	The term $E_{44}$ is

\be\label{p5} E_{44} = {1\over 2} \int\int_{\Omega'\times\Omega'}
V_{LJ}(|{\bf r}-{\bf r'}|) {d^3r\over\Omega_4}{d^3r'\over\Omega_4}\; . \ee

        As above, the integration domain can be separated into several
subdomains. The position dependent part of $E_{44}$ gives a contribution to
$V_{eff}$ which represents the interaction of fictitious $^4$He atoms occupying
the volumes $\Omega_3[{\bf r}_1]$ and $\Omega_3[{\bf r}_2]$

\be\label{p6} E_{44} \leadsto
\int\int_{\Omega_3[{\bf r}_1]\times\Omega_3[{\bf r}_2]}
V_{LJ}(|{\bf r}-{\bf r'}|) {d^3r\over\Omega_4}{d^3r'\over\Omega_4}
\simeq \left(\ds\Omega_3\over\ds\Omega_4\right)^2
V_{LJ}(|{\bf r}_1-{\bf r}_2|) \; . \ee

	As in (\ref{p4}) the last term of Eq. (\ref{p6}) is a long
distance approximation.

	Gathering the contributions (\ref{p2},\ref{p4},\ref{p6}) we see as
stated in the beginning of this section that the long range quasi-particle
interaction is equal to the bare interaction reduced by a factor $\alpha^2$.
Essentially the same result was obtained in \cite{2} using thermodynamical
arguments in momentum space. In the terminology of Ref. \cite{2} the
contribution of $E_{33}+E_{34}$ would correspond to the direct part of the
effective interaction and $E_{44}$ would give the phonon-induced term. Of
course, derivations based on excluded volume arguments such as in \cite{2} or
as presented here (from \cite{17}) are only valid for long wavelengths, {\it
i.e.} for large distances between the $^3$He atoms. Working in real space has
the advantage of providing a simple way to build an effective interaction
sensible also at short distance : if the two atoms get closer one can mimic the
interaction of the dressed particles by introducing a correlation term $g(r)$
describing phenomenologically effects such as the disturbance of the $^4$He
cloud around a $^3$He atom by the other particle. We choose for the correlation
function the following form ($r=|{\bf r}_1-{\bf r}_2|$) :

\be\label{p7} g(r) = \exp \{ -\left( {r_c\over r}\right)^5 \} \; . \ee

	Then the contribution of $E_{34}$ to $V_{eff}$ is approximatively given
by its long range approximation (the r.h.s. of (\ref{p4})) multiplied by a
factor $g(r)$. Similarly the contribution of $E_{44}$ is multiplied by
$g^2(r)$, leading to a total effective interaction

\be\label{p8} V_{eff}(r) = \left[ 1 - (1+\alpha)g(r) \right]^2 V_{LJ}(r)
 \; . \ee

	The short range and long range behavior of the effective interaction
(\ref{p8}) follow the requirements (1) and (2) stated in the beginning  of this
section. Requirement (3) will be fulfilled by a correct choice of the free
parameter $r_c$ in (\ref{p7}). The value $r_c = 3.684$ \AA~ gives the correct
s-wave scattering length  $a=-0.97$ \AA~ (see Ref. \cite{5}). The potential
$V_{eff}$ is shown on Fig. 6 where it is compared to the bare Lennard-Jones
interaction. We have checked that working with a more realistic bare
interaction such as the Aziz potential \cite{18} does not affect the
qualitative picture presented below. Also the potential in films should be
different from the bulk interaction : at the free surface for instance, ripplon
--- and not phonon --- exchange should dominate the long range behaviour. Or
equivalently, the excess volume parameter should be replaced by an ``excess
surface parameter". We will not discuss this effect in the paper.

	In order to test the sensibility of the final results to the effective
interaction we designed another potential $\widetilde{V}_{eff}$ having the
required properties and roughly imitating at long range the potential
derived by Owen using the hypernetted chain approximation \cite{Ow81}.
The potential was chosen to be :

\be\label{p9} \widetilde{V}_{eff} (r) = 4\varepsilon \left[
\left({\ds\sigma\over\ds r}\right)^{12} -\alpha^2
\left({\ds\sigma\over\ds r}\right)^6 - C
\left({\ds\sigma\over\ds r}\right)^7
\cos \left( 2 \pi{\ds r-\sigma \over\ds R}\right) \right] \; . \ee

	This form obviously fulfills requirements (1) and (2) above. The
parameters $C$ and $R$ are chosen so that  $\widetilde{V}_{eff}$ has a zero at
$r=4$ \AA ~ as in Owen's results (this imposes $R=5.776$ \AA) and that the
s-wave scattering length has the correct value (this fixes $C=1.145$). The
corresponding $\widetilde{V}_{eff}$ is plotted on figure 6. Note that the
potential of Ref. \cite{Ow81} is much deeper than $\widetilde{V}_{eff}$ (it has
a minimum at approximatively 19 K) and this would favour dimer's creation. On
the other hand Owen uses no effective mass and this goes against binding. Hence
we cannot directly compare our approach with the one of Ref. \cite{Ow81};
Owen's work is taken here only as an inspiration for designing a new potential
in order to test the sensitivity of our results to the effective $^3$He--$^3$He
interaction.

\section{($^3$He)$_2$ dimers}

	The Hamiltonian describing two $^3$He quasiparticles in the film
has the form:

\be\label{d3} H = {{\bf p}_1^2\over 2M^*} + {{\bf p}_2^2\over 2M^*} +
V_{eff} (|{\bf r}_1 -{\bf r}_2|) + U_{ext}(z_1) + U_{ext}(z_2) \; , \ee

\noindent where $U_{ext}$ is the $^3$He mean field due to both the substrate
and the $^4$He film (such as shown on Figs. 2,3 and 4) and $M^*$ the $^3$He
effective mass for the surface (Andreev) or substrate state ({\it cf} Table 1).
It should be pointed out that the $^3$He particles described by Eq. (\ref{d3})
may, in general, be in two different localized states in the $z$--direction,
{\it i.e.} may belong to two different 2D continua. In this case the effective
masses of both quasiparticles in the Hamiltonian (\ref{d3}) may be quite
different. As explained earlier we restrict ourselves to considering the case
of two $^3$He particles in the same state concerning motion along the
$z$--axis, since  it provides the largest binding energy. In Eq. (\ref{d3}) and
in the following $V_{eff}$ is used as a generic notation valid for both
potentials $V_{eff}$ and $\widetilde{V}_{eff}$. The description might in
general be improved by phenomenologically introducing a $^3$He
concentration--dependent effective mass $M^*$ within the local density
functional approach to $^3$He--$^4$He mixtures \cite{10}. However the
uncertainty in evaluating the dimer binding energy are such that this
correcting term goes beyond the accuracy of the theory (see below).

	We make the following ansatz for the wave function of the two $^3$He
atoms :

\be\label{d4} \Psi({\bf r}_1,{\bf r}_2) = {1\over \ds2\pi}
e^{i {\bf K}_\parallel.{\bf R}_\parallel}
\chi({\bf r}_\parallel) \phi (z_1) \phi (z_2) \; .\ee

	Eq. (\ref{d4}) describes two fermions with opposite spins. ${\bf
R_\parallel}$ is the center of mass coordinate in the ($x,y$) plane. For a
dimer at rest (as we consider in the following) the corresponding momentum
${\bf K_\parallel}$ is zero. The variable ${\bf r_\parallel}$ is the relative
coordinate in the plane. We have separated the ($x,y$) and $z$ direction. The
functions $\phi (z_1)$ and $\phi (z_2)$ describe the motion along the
$z$-axis. We assume (which is exact to the first order of perturbation theory)
that they are of the type shown on Fig. 2, not being affected by the coupling
between the two atoms. So the main point here is that the interaction between
$^3$He atoms just slightly disturbs the motion in the $z$-direction but
entirely changes the relative motion in the plane and leads to a bound state.
In other words a perturbation theory can be applied for describing the normal
motion only but the transverse dynamics of particles should be determined from
the exact Schr\"odinger equation. In practice this means that one has to solve
the 2D Schr\"odinger equation with a potential averaged over the unperturbed
wave functions of $^3$He quasiparticles in the $z$ coordinate. The function
$\phi$ is determined by the solution the equation :

\be\label{d5} \left[ {\ds p_z^2\over \ds 2M^*} + U_{ext}(z) \right]
\phi (z)  = \epsilon \phi (z) \; , \ee

\noindent where $p_z$ is the $z$ component of the momentum ($z = z_1$ or $z_2$)
:

\be\label{d6} {{\bf p}_1^2\over 2M^*} + {{\bf p}_2^2\over 2M^*} =
{p_{z_1}^2\over 2M^*} +{p_{z_2}^2\over 2M^*} +
{{\bf P_\parallel}^2\over 2M} + {{\bf p_\parallel}^2\over 2\mu} \; . \ee

\noindent with $M$ and $\mu$ being respectively the total effective mass
($2M^*$) and the reduced mass ($M^*/2$). Then wri\-ting the Schr\"o\-din\-ger
equa\-tion for the en\-tire wave func\-tion $\Psi$, mul\-ti\-plying by
$\phi^*(z_1)\phi^*(z_2)$ and in\-te\-grating over the variables $z_1$ and
$z_2$, one easily finds :

\be\label{d7} \left[ {\ds {\bf p}_\parallel^2\over \ds 2\mu} + \langle V_{eff}
({\bf r_\parallel}) \rangle \right] \chi ({\bf r_\parallel}) =
E_{dim} \chi ({\bf r_\parallel}) \; , \ee

\noindent where

\be\label{d8} \langle V_{eff} ({\bf r_\parallel}) \rangle =
\int \phi^2(z_1) \phi^2(z_2) V_{eff} ({\bf r}_1 - {\bf r}_2) dz_1 dz_2
\; . \ee

	According to our perturbation scheme, all the $z$-dependence has been
removed from Eq. (\ref{d7}). The substrate and the $^4$He density play an
indirect role, through the determination of $\mu$ ({\it i.e.} $M^*$) and
the $z$-wave function.

	It appears that for all the cases we are interested in (the Andreev
state or the substrate state) the $z$-wave function can be represented to a
fairly good approximation by a simple gaussian (characterized by its
half-width). In order to explore the sensitivity of the dimer binding
energy to the two most relevant parameters, namely the effective $M^*$
and the width of the $^3$He wave function, rather than considering only the
substrates listed in Table 1, we will give results for various
values of $M^*$ and normalized gaussians for $\phi^2(z)$ with various
half width.

	In the limit of a large half-width, although common perturbation theory
cannot be directly applied when looking for the solution of the Schr\"odinger
equation, one can use the Fermi renormalization technique (see
\cite{Lif80} and appendix A) to obtain the binding energy with a logarithmic
accuracy:

\be\label{d9}
|E_{dim}| \simeq {\ds \hbar^2\over \ds M^* r_0^2} \exp \left(
- {\ds M^*_b\over \ds M^*} {\ds L \over\ds |a|} \right) \; ,
\ee

\noindent where $r_0$ is a quantity of the order of the interaction potential
range, $M^*_b$ is the effective mass of a $^3$He quasiparticle in
bulk $^4$He, and $L$ is defined as:

\be\label{d10} {\ds 1 \over \ds L} = \int |\phi(z)|^4 dz \; . \ee

	Here $L$ is of the order of the half-width, $w$, and it is a measure of
the spatial extension of the $z$-wave function. The result above is valid only
for large values of $L/|a|$ (see Appendix). One can see from Eq. (\ref{d9})
that in this limit the requirements of Sec. 3 are enough to determine the order
of magnitude of the dimer binding energy.

	In the opposite limit of very localized states --- such as those we are
primarily interested in --- our numerical results show a great sensibility of
the binding energy to the details of the potential (the results for $E_{dim}$
as a function of the half-width for different effective masses are shown on
Fig. 7 for $V_{eff}$ and on Fig. 8 for $\widetilde{V}_{eff}$). For instance,
for  $M^*/m_3 \simeq 3.1$ and $w\simeq 0.5$ \AA~ the binding energies
estimated with $V_{eff}$ and $\widetilde{V}_{eff}$ differ from each other by
two orders of magnitude. Under these conditions an accurate quantitative
prediction would be illusive. We just note that if $V_{eff}$ mimics
accurately the exact quasiparticle interaction, it would be very difficult to
observe the formation of ($^3$He)$_2$ bound states in experiment. On the other
hand, if the effective potential $\widetilde{V}_{eff}$ can be applied in the
case in question, then for a magnesium substrate ($M^*/m_3 = 2.9$ and  $w=0.9$
\AA) the dimer binding energy would be $E_{dim} = -1.1 $ mK, which is a
reachable temperature with modern experimental techniques. A problem might
occur though because the first $^4$He layer on a Mg substrate could be solid
(see the discussion in Sec. 2). The next candidate for dimer formation would
then be hydrogen. Using the two potentials designed in Sec. 3 we could not get
a reasonable dimer binding energy, mainly because  $M^*$ in this case is not
large enough. However, other pseudo-potentials might give a different result.

	Note also that the dimer binding energies are very sensitive to the
exact value of the s-wave scattering length. This can be seen in the large $L$
(or equivalently large $w$) limit from Eq. (\ref{d9}). We verified numerically
that this is also true for small $L$ : taking the outdated value $a=-1.5$ \AA~
we obtained (using $\tilde{V}_{eff}$) a dimer binding energy on magnesium
$E_{dim}=-6.3$ mK. Hence experimental information on dimer binding could give
valuable insight on the exact value of $a$.

	Although not quantitatively predictive, our study allows us
nevertheless to draw a very clear qualitative picture. In highly compressed
substrate layers the $^3$He effective mass is large. This reduces the kinetic
energy and favours the creation of ($^3$He)$_2$ bound states. The small spatial
extension of the wave function in these layers is also favorable to the
formation of dimers as can be seen from Eq. (\ref{d9}) and Fig. 7 and 8. From
these figures one can also see that there is an optimal width of the order of
0.5 \AA. It is interesting to note that the more attractive the substrate, the
closer one approaches this value (see Table 1). The proper way to observe
dimers would then be to choose a substrate for which the first $^4$He layer in
close to be solid but still remains liquid.

\section{Concluding remarks}

    	After having classified various substrates according to their ability
to solidify one or two layers of a multilayer helium film, we have concentrated
on cases in which helium remains liquid. In addition to the already known
Andreev states we have shown that $^3$He impurities were able to form a new 2D
Fermi system near the substrate. After designing a sensible schematic
interaction we have computed the binding energy of ($^3$He)$_2$ dimers in the
limit of small $^3$He concentration. We found that for attractive substrates
(such as magnesium or hydrogen) there was a reasonable hope to form dimers with
sizeable binding energy. Note here that the binding energy  of dimers in
Andreev surface states is extremely small according to all our estimates, so
the eventual observation of ($^3$He)$_2$ dimers would be a direct consequence
of the existence of $^3$He substrate states.

	In this paper the quantitative calculation of the dimer binding energy
was based on a semi--empirical effective interaction and the one--particle wave
functions obtained within a density--functional approach. It would be very
useful to carry out the appropriate calculations with the help of other
theoretical approaches ({\it e.g.} Ref. \cite{kro}). This would allow a
better understanding of the accuracy of the present computations.

	We hope that these results will motivate experimental study of the
substrate $^3$He states as was proposed earlier in \cite{11,11b}. The
eventual formation of ($^3$He)$_2$ dimers would be a very interesting
consequence of the existence of these states leading to an amazingly rich phase
diagram. An experimentalist could face an extra Kosterlitz--Thouless phase
transition, liquefaction of ($^3$He)$_2$ or polymerization of $^3$He, crossover
from a Bose gas of dimers to a 2D Fermi fluid (with strong pair correlation)
upon increasing the $^3$He concentration, {\it etc.}
\cite{7,7b,Leg80,miy,Noz85,ost}. The dimer binding energy is a {\it
cornerstone} characteristic of all these phenomena which are exciting
objectives for experimental and theoretical studies.

	A natural continuation of the present work would be to study the
binding of two dimensional clusters of $^3$He atoms. Although Fermi statistics
favours droplets with even numbers of $^3$He atoms, there might be no upper
limit for $N$, and our results could be a hint of the existence of a 2D liquid
phase of $^3$He. On the basis of a regular arrangement in two dimensions (with
6 nearest neighbours per atom) we estimate the saturation energy of the liquid
phase to be (in the most favourable case) of the order of 5 to 6 mK. This is to
be related to the recent finding of Brami {\it et al.} \cite{Cla94} whose
variational results lead to propose at low temperature a new ``self-condensed"
fluid phase for pure $^3$He films on graphite. As the presence of $^4$He
favours the formation of $^3$He dimers it could also favour the formation of a
liquid phase.

\

{\bf Acknowledgements:} We thank M. W. Cole and M. Devaux for fruitful
discussions. This research was supported in part by the Deutsche
Forschungsgemeinschaft (BA 1229/4--1). One of the authors (E. B.) greatly
appreciates the warm hospitality extended to him at the Pennsylvania State
University where part of this work was done under the NSF Grant DMR--9022681.

\appendix

\section{}
\setcounter{equation}{0}

	In this appendix we derive the expression (\ref{d9}) for the dimer
binding energy $E_{dim}$ using the Fermi renormalization technique (see
\cite{Lif80}). According to this method we introduce a weak two-particle
pseudo-potential $V_f(r)$ which is assumed to meet the perturbation theory
criterion. Let us emphasize that such an pseudo-potential has nothing to do
with the interaction between quasiparticles which, indeed, cannot be treated in
terms of a perturbation theory at all.

	The pseudo-potential $V_f$ is supposed to result in the
true scattering amplitude when calculated within the Born approximation :

\be\label{a1} a = {\ds M^*_b \over \ds 4\pi\hbar^2} \int V_f(r) d^3r \; , \ee

\noindent where $M^*_b$ is the effective mass of a $^3$He quasiparticle in bulk
$^4$He ($M*_b=2.38 m_3$). The idea of the method is to carry out all
calculations in the framework of a perturbation theory for $V_f$ and then to
express the final formulae through the true s-wave scattering length $a$ only
by means of the renormalization (\ref{a1}). Thus the method works while all
obtained expressions contain the pseudopotential $V_f$ only in the integral
form (\ref{a1}).

	The solution of the Schr\"odinger equation, {\it i.e.} of Eq.
(\ref{d7}) with $V_f$ replacing $V_{eff}$, may
be expressed in the form (see {\it e.g.} Ref. \cite{8,Eco83})

\be\label{a2}
E_{dim} \simeq -{\ds \hbar^2\over \ds 2\mu r_0^2}
\exp \left[ -{\ds\hbar^2\over\ds\mu}
\left| \int_0^\infty \langle V_f \rangle\rho d\rho \right|^{-1} \right] \; ,
\ee

\noindent where $r_0\sim |a|$ is of the order of the interaction range and
$\langle V_f \rangle$ is defined by (\ref{d8}) with $V_f$ replacing $V_{eff}$.
Let us recall once again that the pseudopotential $V_f$ is assumed to meet the
criterion of a perturbation theory : for instance the integral entering Eq.
(\ref{a2}) should converge. Denoting this integral as $I$ one can write :

\be\label{a3} I = \int_0^\infty \langle V_f \rangle\rho d\rho
 = \int V_f(\sqrt{ \rho^2 + z^2})  F(z) \rho d\rho dz \; , \ee

\noindent where

\be\label{a4} F(z) = \int |\phi(t-z/2)|^2\; |\phi(t+z/2)|^2 dt \; . \ee

	Let the quantity $w$ be a characteristic localization range in the
$z$-direction for the wave function $\phi$ ($w$ describes the half-width
introduced in section 4). If the localization length is sufficiently large, $w
\gg r_0$, one can easily find that

\be\label{a5} I = {\ds 1\over \ds 2\pi L} \int V_f(r) d^3r \quad \mbox{with}
\quad {\ds 1\over \ds L} \equiv F(0) = \int |\phi(t)|^4 dt \; . \ee

	Evidently $L$ is of the same order as $w$ (if $\phi^2$ is a
normalized gaussian of half-width $w$, then $L=w \sqrt{\pi /\ln 4}$).
Combining Eq. (\ref{a5}) and the renormalization relation (\ref{a1}) yields :

\be\label{a6} E_{dim} \simeq - {\ds \hbar^2 \over \ds M^* r_0^2}
\exp \left[ -{\ds M^*_b\over M^*} {\ds L\over \ds |a|} \right] \; . \ee

	In (\ref{a6}) we have replaced the reduced mass by its value
{\it for the state $\phi(z)$ considered} :
$\mu = M^*/2$. We see that the binding energy is expressed in terms of the true
scattering amplitude and does not contain a pseudopotential in any explicit
form. Thus using the Fermi method is completely justified in the limiting case
$w \gg r_0$. Note, however, that it is difficult to exactly determine
the quantity $r_0$ (see \cite{Eco83}) which could, in general,
be pseudo-potential dependent. On the other hand,
its order of magnitude is known ($r_0 \sim |a|$) and, in fact, it does not
enter the exponential factor which determines the dominant behavior of
$E_{dim}$.

	In the limiting case $w \ll r_0$ the integral $I$ obviously reduces to

\be\label{a7} I = \int V_f (\rho) \rho d\rho \; , \ee

\noindent and the binding energy explicitly depends on what kind of a
pseudopotential is chosen. It means that the renormalization method is no
longer valid and the 2D Schr\"odinger equation with the real interaction
potential should be solved to find the binding energy. In this paper we have
chosen to perform the calculation of $E_{dim}$ by means of semi--empiric
pseudopotentials which gave rise to reasonable estimates for quantities
measured in experiment (see Sec. 2).

\newpage

{\center {\Large {\bf Table captions }}}
\vspace{1cm}

{\bf Table 1.} ~Energy ($\epsilon$) , effective mass ($M^*$) and half-width
($w$) of the $^3$He impurity state on various substrates and for the Andreev
state. In the case of H$_2$, the value $C_3=$ 1000 K\AA$^3$ extracted from
different experiments \cite{14b} is different from the theoretical
value of 360 K\AA$^3$ \cite{14}.

\vspace{0.5cm}
\begin{center}
\begin{large}
\begin{tabular}{|cccccc|}\hline\hline
 & & & & & \\
 substrate & $C_3$ [ K\AA$^3$ ] & $D$ [ K ] & $\epsilon$ [ K ] &
$M^*/m_3$
& $w$ [ \AA ~] \\
 & & & & & \\
 & & & & & \\ \hline
 & & & & & \\
   Cs       &  673  &  4.4  &  $-4.74 $   &  1.51 &  2.0 \\
 & & & & & \\
   Rb       &  754  &  5.0  &  $-4.23 $   &  1.78 &  1.9  \\
 & & & & & \\
   K        &  812  &  6.3  &  $-4.15 $   &  1.84 &  1.8  \\
 & & & & & \\
   Na       &  1070 & 10.4  &  $-4.04 $   &  2.02 &  1.4  \\
 & & & & & \\
   Li       &  1360 & 17.1  &  $-4.17 $   &  2.28 &  1.2  \\
 & & & & & \\
   Mg       &  1780 & 32.0  &  $-5.00 $   &  2.90 &  0.9  \\
 & & & & & \\
   H$_2$    &   360 & 28.0  &  $-5.51 $   &  2.41 &  0.9  \\
            &  1000 & 28.0  &  $-5.00 $   &  2.61 &  0.9  \\
 & & & & & \\
 $^4$He free surface &   &  &  $-5.20 $   &  1.35 & 3.6  \\
 & & & & & \\ \hline\hline
\end{tabular}
\end{large}
\end{center}

\newpage
{\center {\Large {\bf Figure captions }}}
\vspace{1cm}

{\bf Figure 1.} ~Phase diagram of $^4$He on various substrates at zero
temperature (form \cite{13}). The dotted line is the wetting line
\cite{12}.

\vspace{1cm}

{\bf Figure 2.} ~$^4$He liquid density (dotted line), $^3$He substrate state
wave-function (with arbitrary units) and $^3$He average field on a Cs substrate
(less attractive) and Li substrate (more attractive).

\vspace{1cm}

{\bf Figure 3.} ~$^4$He film (0.6 atoms per \AA$^2$, {\it i.e.} 7.7 atomic
layers) and $^3$He mean field on a magnesium substrate. The solid line is the
Density Functional mean field $U_{ext}$ and the dashed line is the result of
Lekner's approach $U^L_{ext}$.

\vspace{1cm}

{\bf Figure 4.} ~Infinite $^4$He film and $^3$He mean field on a cesium
substrate. The solid line is the Density Functional mean field $U_{ext}$  and
the dashed line is the result of Lekner's approach $U^L_{ext}$.

\vspace{1cm}

{\bf Figure 5.} ~Schematic representation of the behaviour of two $^3$He atoms
situated at points ${\bf r}_1$ and ${\bf r}_2$ in bulk liquid $^4$He. Atom $i$
$(i=1,2)$ occupies a volume $\Omega_3[{\bf r}_i]$ of which the $^4$He's are
expelled (see the text). The shaded zone represents the volume occupied by
liquid $^4$He.

\vspace{1cm}

{\bf Figure 6.} ~$^3$He interaction potentials. The solid line represents the
bare Lennard-Jones potential $V_{LJ}$, the dashed lines represent the effective
potentials $V_{eff}$ (long dashes) and $\widetilde{V}_{eff}$ (short dashes).

\vspace{1cm}

{\bf Figure 7.} ~$E_{dim}$ as a function of $w$ computed with $V_{eff}$ for the
effective masses $M^*/m_3 =$ 3.1, 3.15 and 3.2.

\vspace{1cm}

{\bf Figure 8.} ~$E_{dim}$ as a function of $w$ computed with
$\widetilde{V}_{eff}$ for the effective masses $M^*/m_3 =$ 2.8, 2.9 and 2.95.
Note the change of energy scale with respect to Fig. 7.

\end{document}